\documentclass[12pt]{article}

\def\half{\textstyle{\frac{1}{2}}}

\def\H{{\cal H}}

\def\p{\varphi}

\def\H{{\cal H}}

\def\l{\lambda}

\def\ra{\rightarrow}
\def\tint{{\textstyle\int}}
\def\hg{{\hat g}}
\def\hp{{\hat\pi}}

\def\b{\begin{eqnarray*}}  
\def\e{\end{eqnarray*}}    
\def\bn{\begin{eqnarray}}  
\def\en{\end{eqnarray}}   

\def\<{\langle}
\def\>{\rangle}
\def\k{\kappa}

\def\bk{\mathbf k}

\def\no{\nonumber}

\def\ds{d^s\!x}
\def\k{\kappa}

\def\hk{\hat{\kappa}}

\def\hp{\hat{\pi}}
\def\{{\lbrace}
\def\hv{\hat{\varphi}}
\def\}{\rbrace}

\begin{document}

\title{Quantum Field Theory Deserves Extra Help}  
  \author{John R. Klauder\footnote{klauder@ufl.edu} \\
Department of Physics and Department of Mathematics  \\
University of Florida,   
Gainesville, FL 32611-8440}
\date{ }

\maketitle 

\begin{abstract}
Today's quantum field theory (QFT) relies heavenly on canonical quantization (CQ), which fails for
$\p^4_4$ leading only to a ``free'' result. Affine quantization (AQ), an alternative quantization 
procedure, leads to a ``non-free'' result for the same model. Perhaps adding AQ to CQ can improve 
the quantization of a wide class of problems in QFT. \end{abstract}

\section{What is AQ?}
The simplest way to understand AQ is to derive it from CQ. The classical variables, $p\,\&\,q$, lead to self-adjoint quantum operators, $P\,\&\,Q$, that cover the real line, i.e., $-\infty< P, Q<\infty$, and obey
$[Q, P]\equiv QP-PQ= i\hbar 1\!\!1$. Next we introduce several versions of $Q\,[Q, P] = i\hbar\,Q$, 
specifically
\bn &&\{Q [Q,P] + [ Q,P] Q\}/2 = \{Q^2P-QPQ + QPQ - PQ^2 \}/2 \no \\
    &&\hskip3em = \{Q(QP+PQ) - (QP+PQ)Q\}/2= [Q,QP+PQ]/2\;. \en                                
This equation serves to introduce the `dilation' operator $ D \equiv (QP+PQ)/2$\footnote{Even if $Q$ does not cover the whole real line, which means that $P^\dag\neq P$, yet $P^\dag Q=P\,Q$. This leads to $D=(QP+P^\dag Q)/2 =D^\dag$.} which leads to
$[Q,D] = i\hbar\, Q$. While $P (=P^\dag)\, \&\,Q (=Q^\dag)$ are the foundation of CQ, $D (=D^\dag)\,\&\, Q (=Q^\dag)$ are the foundation of AQ. Another way to examine this story is to let $p, q \ra P, Q$, while $d\equiv pq, q \ra D, Q$. 

Observe, for CQ, that while $q\,\&\,Q$ range over the whole real line, that is not possible for AQ.
If $q \neq0$ then $d$ covers the real line, but if $q=0$ then $d=0$ and $p$ is helpless. To eliminate this possibility
we require $q\neq 0\,\&\,Q\neq 0$. While this may seem to be a problem, it can be very useful to limit such variables, like
$0<q\,\&\,Q<\infty$, or $-\infty<q\,\&\,Q<0$, or even both.\footnote{For example, affine quantization of gravity can 
restrict operator metrics to positivity, i.e., $ \hg_{ab}(x)\,dx^a\,dx^^b>0$, straight away \cite{11}.}

\section{A Look at Quantum Field Theory}
\subsection{Selected poor and good results}
Classical field theory normally deals with a field $\p(x)$ and a momentum $\pi(x)$, where $x$ denotes a spatial point in an underlying 
space.\footnote{In order to avoid problems with spacial infinity we restrict our space to the surface of a large, $(s+1)$-dimensional sphere.}

A common model for the Hamiltonian is given by
  \bn H(\pi,\p)=\tint \{ \half[ \pi(x)^2 +(\overrightarrow{\nabla}(x))^2 +m^2 \,\p(x)^2] \label{r}
  + g\,\p(x)^r\}\;d^s\!x \;, \en
  where $r \geq 2$ is the power of the interaction term,  $s\geq 3$ is the dimension 
  of the spatial field, and $n=s+1$, which adds the time dimension. Using CQ, such a model is nonrenormalizable when $r>2n/(n-2)$,
  which leads to ``free'' model results \cite{5}. Such results are even true  for $r=4$ and $n=4$, which is a  case where $r=2n/(n-2)$ \cite{2,3,4}. When using AQ, the same models lead to ``non-free'' results \cite{5,6}.
  
 Solubility of classical models involves only a single path, while quantization involves a vast number of paths, a fact well illustrated by path-integral quantization. The set of acceptable paths can shrink significantly when a nonrenormalizable term is introduced. Divergent paths are like those for which $\p(x,t)= 1/z(x,t)$ when  $z(x,t)=0$. A procedure that forbids possibly divergent paths would eliminate nonrenormalizable behavior. As we note below, AQ provides such a procedure.

\subsection{The classical and quantum affine story}
Classical affine field variables are $\k(x)\equiv \pi(x)\,\p(x)$ and $\p(x)\neq0$. The quantum versions are $\hk(x)\equiv[\hv(x)\hp(x)+\hp(x)\hv(x)]/2 $ and $\hv(x)\neq0$, with $ [\hv(x), \hk(y)]=i\hbar\,\delta^s(x-y)\,\hv(x)$. The affine quantum version of  (\ref{r}) becomes 
\bn &&\H(\hk,\hv)=\tint\{\half[\hk(x)\,\hv(x)^{-2}\,\hk(x) +(\overrightarrow{\nabla}\hv(x))^2  + m^2\,\hv(x)^2] \no \\
&&\hskip12em + g\,\hv(x)^r\} \; d^s\!x \;. \en
The spacial differential term restricts $\hv(x) $   to continuous operator functions, maintaining 
$\hv(x)\neq0$. In that case, it follows that $\hv(x)^{-2}>0$ which implies that $|\hv(x)|^r<\infty$ for all $r<\infty$, a most remarkable feature!\footnote{Concern for the term $\hv(x)^{-2}\neq 0$ has been resolved by successful usage 
of $[\hv(x)^2+\epsilon]^{-1}$, where $\epsilon =10^{-10}$ \cite{5, 6}.}

Adopting a Schr\"odinger representation, where $\hv(x)\ra \p(x)$, simplifies  $\hk(x)\,\p(x)^{-1/2} =0$, which also implies that $\hk(x)\,\Pi_y\,\p(y)^{-1/2}=0$. This relation suggests that a general wave function is like
$\Psi(\p)= W(\p) \,\Pi_y\,\p(y)^{-1/2}$, as if $\Pi_y\,\p(y)^{-1/2}$ acts as the representation 
of a family of similar wave functions.

We now take a Fourier transformation of the absolute square of a regularized wave
    function that looks like\footnote{The remainder of this article updates and improves a recent article by the author \cite{88}.}
  \bn F(f)=\Pi_\bk \,\tint \{ e^{if_\bk \p_\bk} |w(\p_\bk)|^2 (ba^s) \,|\p_\bk|^{-(1-2ba^s)}\,
 d\p_\bk\}\;.\label{6} \en
       Normalization ensures that if all $f_\bk =0$, then $F(0)=1$, which leads to
       \bn F(f)=\Pi_\bk \tint \{ 1- \tint(1- e^{if_\bk \p_\bk}) |w(\p_\bk)|^2 (ba^s)\,d\p_\bk/
       |\p_\bk|^{(1- 2ba^s)}\,\}\;.  \en
      Finally, let $a\ra 0$ to secure the normal Fourier transformation\footnote{Any change of $w(\p)$ due to $a\ra0$ is left implicit here.}
 \bn F(f)= \exp\{ -b\tint \ds(1- e^{if(x)\,\p(x)}) |w(\p(x)|^2 d\p(x)/|\p(x)|\,\} \;. \label{2} \en
 
 This particular process side-steps any divergences that may normally arise in  $|w(\p(x)|$ when using
 more traditional procedures.
  
\section{The Absence of Nonrenormalizablity, and The Next  Fourier Transformation}
  Observe the factor $|\p_\bk|^{-(1-2ba^s}$ in (\ref{6}) which is prepared to insert a
  zero divergence for each and every $\p_\bk$ when $a\ra0$. However, the factor $ba^s$ in (\ref{6}) turns that possibility into a very different story given  in (\ref{2}).
  
  Another Fourier transformation can take us back to suitable functions of the field $\p(x)$. That task involves pure mathematics, and it deserves a separate analysis of its own.

\end{document}